\documentclass[conference]{IEEEtran}
\usepackage{xcolor}
\usepackage{multirow}

\ifCLASSINFOpdf
\usepackage[pdftex]{graphicx, adjustbox}
\else
\fi
\begin{document}
%
\title{An Implementation for Dynamic Application Allocation in Shared Sensor Networks}

\author{\IEEEauthorblockN{Carmen~Delgado,~Sergio~Batista,\\~Mar\'ia~Canales, and~Jos\'e Ram\'on~G\'allego}
\IEEEauthorblockA{Arag\'on Institute of Engineering Research\\
Universidad de Zaragoza\\
Zaragoza, Spain\\
Email: cdelga@unizar.es}
\and
\IEEEauthorblockN{Jorge~Ort\'in,}
\IEEEauthorblockA{Centro Universitario de la Defensa and \\
Arag\'on Institute of Engineering Research\\
Universidad de Zaragoza\\
Zaragoza, Spain\\
Email: jortin@unizar.es}
\and
\IEEEauthorblockN{Matteo~Cesana}
\IEEEauthorblockA{Dipartimento di Elettronica,\\Informazione e Bioingegneria\\ Politecnico di Milano\\Milano, Italy\\
Email: matteo.cesana@polimi.it}}

%


\maketitle

\begin{abstract}
We present a system architecture implementation to perform dynamic application allocation in shared sensor networks, where highly integrated wireless sensor systems are used to support multiple applications. The architecture is based on a central controller that collects the received data from the sensor nodes, dynamically decides which applications must be simultaneously deployed in each node and, accordingly, over-the-air reprograms the sensor nodes. Waspmote devices are used as sensor nodes that communicate with the controller using ZigBee protocol. Experimental results show the viability of the proposal. \end{abstract}

%
\IEEEpeerreviewmaketitle

\section{Introduction}
\IEEEPARstart{W}{ireless} Sensor Networks (WSNs) are one of the key enabling building blocks for the Internet of Things. Looking back at the evolution of WSNs, a clear trend can be observed moving from stand-alone, application-specific deployments to the emergence of \emph{shared sensor networks}, where highly integrated wireless sensor systems are used to support multiple services and applications in deployment domains such as Smart Cities, Smart Home and Buildings and Intelligent Transportation Systems. 

The aforementioned trend calls for novel design good practices  to overcome the limits in flexibility, efficiency and manageability of vertical, task-oriented and domain-specific  WSNs. Being the research field that recent, a common terminology is still missing and the technical papers often use different wording for similar concepts; as an example,  \emph{shared sensor networks}, \emph{virtual sensor networks} and \emph{multi-application sensor networks} are often used almost interchangeably in the literature. The interested reader may refer to the following surveys on the topic \cite{khan2015, Farias2016}. In this work, we will use the term \emph{shared sensor network} (SSN) to define a physical sensor network infrastructure which can be used to support multiple concurrent applications and services, and where the ownership of the physical infrastructure is decoupled from the ownership of the applications and services. Generally speaking, the efficient realization of SSNs requires technologies and solutions in different domains ranging from the node level, where the sensor nodes must be able to support and run applications in a transparent way, up to the network level, where effective platforms and solutions are required to manage and reconfigure on-the-fly the network resources.   

At the node level, the design of abstraction layers and primitives on a single sensor node to overcome the problem of application-platform dependency has been already addressed. As an example, Mate\'e \cite{Levis2002}, ASVM \cite{Levis2005}, Melete \cite{Yu2006} and VMStar \cite{Koshy2005} are frameworks for building application-specific virtual machines over constrained sensor platforms. At the network level there are two main building blocks which are usually tightly coupled: (i) management platforms to support multiple application sharing a common physical infrastructure, and (ii) tools/algorithms to allocate the physical resources to the multiple applications. Representative examples of management platforms are SenSHare \cite{Leontiadis2012} and UMADE \cite{5465985}, which create multiple overlay sensor networks which are ``owned'' by different applications on top of a shared physical infrastructure. As far as resource allocation in SSNs is concerned, the authors of \cite{Xu2010} focus on environmental monitoring applications and propose a centralized optimization framework which allocates applications to sensor nodes minimizing the variance in sensor readings. Ajmal \emph{et al.} \cite{Ajmal2014} propose a decision algorithm to dynamically ``admit'' applications to a physical sensor network infrastructure. The research streamline on \emph{sensor mission assignment} \cite{Porta2014, Edalat2016} also addresses a resource allocation problem in WSNs; namely, the problem is to jointly allocate the physical sensor resources to incoming applications while incorporating admission control policies. In our past work \cite{Delgado18}, we extended those works considering networking-related aspects, including the possibility to re-configure the sensor network by moving applications which were previously deployed, and further modelling situations where multiple applications can be concurrently deployed at a sensor node.

While our work in \cite{Delgado18} focuses on a mathematical programming framework to model the joint problem of application/service admission control and  resource allocation to the admitted applications, in this work we present a system architecture and implementation that provides the basis to deploy those features in an actual SSN. The remaining manuscript is organized as follows: Section \ref{sec:system} describes the system architecture and its implementation (communications, sensor nodes, applications, central controller). In Section \ref{sec:results} experimental results validating the implementation are shown and concluding remarks are finally reported in Section \ref{sec:conclusions}.

\section{System Architecture and Implementation}
\label{sec:system}

The proposed system architecture aims to manage a shared physical WSN over time. As stated above, 
we look at the reference scenario where a single SSN infrastructure provider (SSN-IP) owns an infrastructure which can be accessed by multiple application providers which issue requests to deploy specific applications/services.

\begin{figure}[!t]
\centering
\includegraphics[width=2.8in]{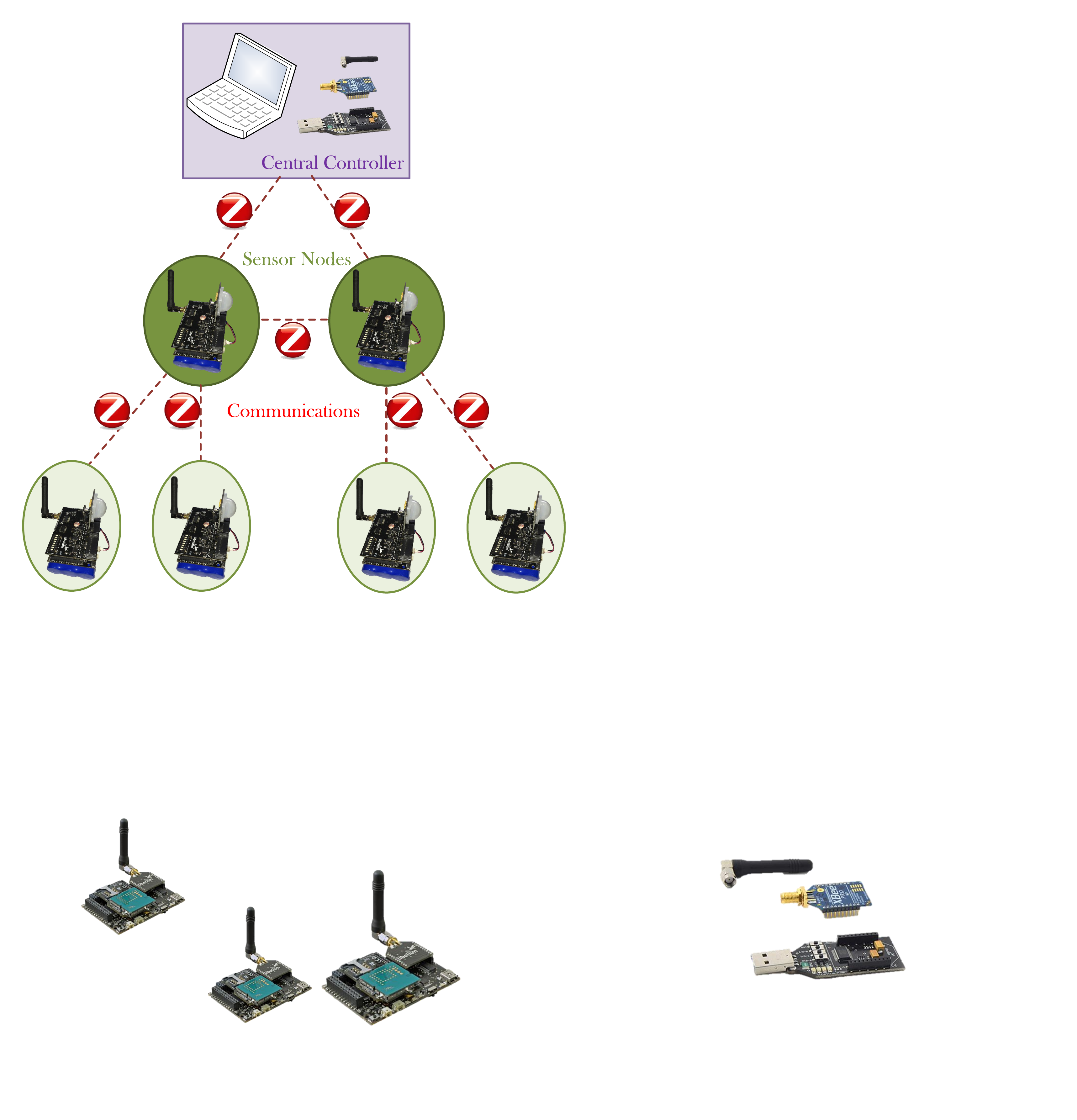}
\caption{Network topology scenario}
\label{fig_scenary}
\end{figure}

In the aforementioned scenario, different applications will be running in the sensor nodes. Each of these nodes will send the monitored data through multihop wireless communications to a sink node, which acts as a central controller. This controller not only collects the received data from the sensor nodes, but dynamically decides which applications must be deployed in each node and, accordingly, reprograms the sensor nodes. Figure~\ref{fig_scenary} schematically shows this architecture composed by three main elements: central controller, sensor nodes and communications. 
In the following we describe in detail these elements and their practical implementation.

\subsection{Communications}
\label{System_communications}

ZigBee protocol \cite{ZigBee} has been chosen to implement wireless multihop communications between the sensor nodes and the central controller. ZigBee is an IEEE 802.15.4-based standard characterized by high reliability, low cost, low power, scalability and low data rate (up to 250 kbps), which is commonly used in WSN implementations \cite{alhmiedat2015}. More specifically, XBee ZigBee-Pro S2 modules \cite{XBeeProUser} have been used. 

It must be noted that in a ZigBee network there are three different device types: Coordinator, Router and End device:

\begin{itemize}
\item \textbf{Coordinator}: each ZigBee network must have one. It selects the channel and PAN ID to start the network, allows routers and end devices to join the network, assists in routing data and has to be always awake.
\item \textbf{Router}: It must join a ZigBee network before it can transmit, receive or route data. After that, it can allow routers and end devices to join the network and route data. It has to be always awake.
\item \textbf{End device}: It must join a ZigBee network before it can transmit or receive data. To do so, it must be associated to a router or a coordinator (its parent). It must always transmit and receive RF data through its parent and can neither allow devices to join the network nor route data. On the contrary, it can sleep.
\end{itemize}

In our implementation, the central controller acts as the coordinator, creating and managing the network whereas the sensor nodes are configured either as routers or end devices, depending on the network requirements and topology. As stated above, routers and coordinator can route data and are interconnected forming a mesh topology with multihop routing. However, each end device is only associated to a parent node (router or coordinator). This can be seen in Figure~\ref{fig_scenary}, where the coordinator and all the routers (darker green) within transmission range can directly communicate with each other while the end devices can only communicate with their parents. In the end, the deployed topology enables a bidirectional communication between all the devices. 

\begin{figure}
\centering
  \begin{minipage}{0.22\textwidth}
    \centering
    \includegraphics[width=0.85\textwidth]{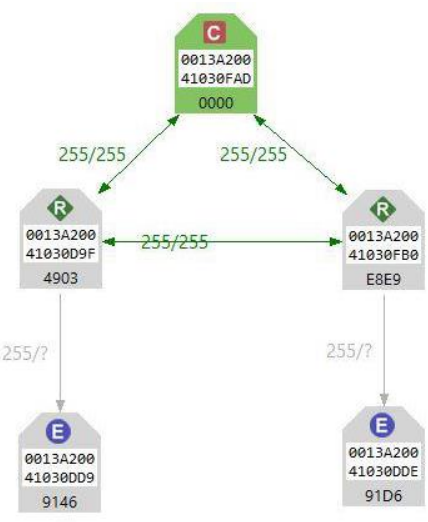}\\
    {\small (a)} 
  \end{minipage}%
  \hspace{5mm}
  \begin{minipage}{0.22\textwidth}
    \centering
    \includegraphics[width=0.95\textwidth]{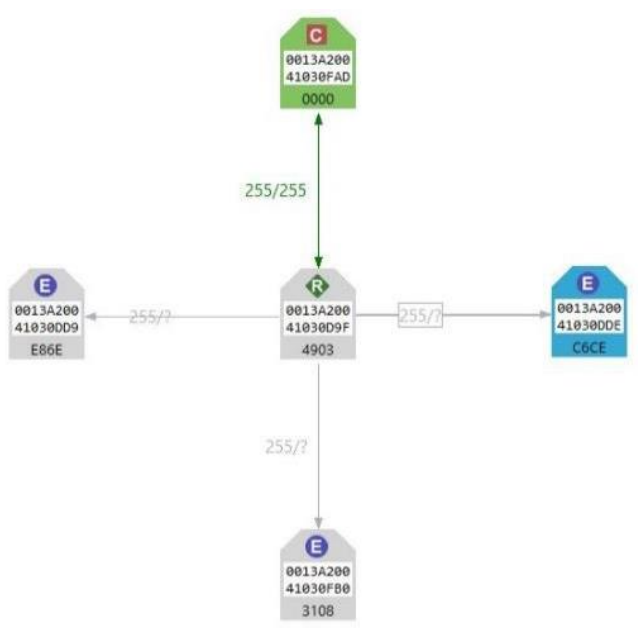}\\
    {\small (b)} 
  \end{minipage}
  \caption{Network topology examples}
  \label{fig_schemas}
\end{figure}

The XBee modules for each type of device have been configured with the XCTU software \cite{XCTU}. Figure~\ref{fig_schemas} shows some of the configured reference scenarios. As previously explained, only end devices are allowed to sleep in a ZigBee network. Thus they are configured in sleep mode and wake up periodically to send the gathered information or to check if they have anything to receive, as it is detailed in section~\ref{architec_sensor}. If an end device is sleeping when its parent has something to send to it, the router keeps the data in its buffer until the end device wakes up and asks if there is any message for it. 

\begin{figure}[!t]
\centering
\includegraphics[width=3.1in]{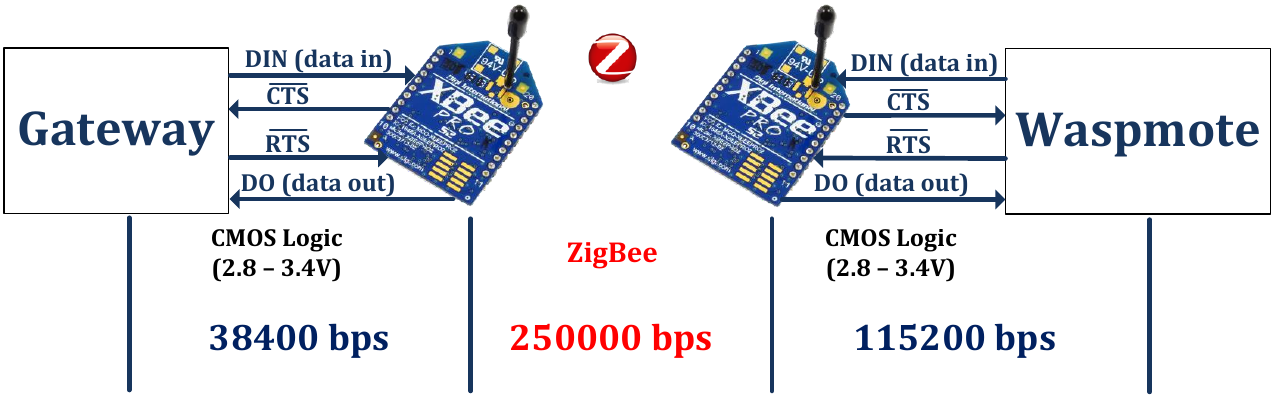}
\caption{Serial and wireless communications between XBee-based devices}
\label{fig_link}
\end{figure}

Finally, it is important to note that the communication between the XBee RF Modules and the host device (sensor node or central controller) is done through a logical-level asynchronous serial port \cite{XBeeProUser}. The maximum transmission rate for this serial communication is 115200 bps, being the value configured for the sensor nodes in our implementation. For the coordinator, the value has been configured to 38400 bps, as it is explained in section~\ref{architec_central}. Therefore, as shown in Figure~\ref{fig_link}, even in a single-hop wireless link between two devices, there are three communications links, which can limit the transmission data rate of a ZigBee application.
 

\subsection{Sensor nodes}
\label{architec_sensor}

\subsubsection{Hardware}
\label{hardware}

The devices used as sensor nodes are Waspmotes Pro (v1.2) developed by Libelium \cite{Libelium}. Waspmote is an open source wireless sensor platform specially focused on the implementation of low consumption nodes that are completely autonomous and battery powered, offering a variable lifetime between 1 and 5 years depending on the duty cycle and the radio used. They have an ATmega 1281 processor at 14.7456 MHz, SRAM of 8 KB, EEPROM  of 8KB and Flash of 128 KB. They also have a micro SD slot with a 2GB SD card. In addition, they are equipped with a Real Time Clock (RTC) that controls the low consumption modes, and a battery connector with a rechargeable Lithium Ion battery of 6600~mAh. The Waspmotes are connected to the XBee ZigBee-Pro S2 module through Socket 0 and to an Events Sensor Board v2.0 \cite{WaspEventsSB} through the I/O Sensor pins. This board has 8 sockets to connect different sensors, where Socket 7 is used for digital signals and the others for analogical ones. Four different sensors have been used in our implementation: the MCP9700A Temperature Sensor (connected to Socket 5), the 808H5V5 Humidity Sensor 3.3V (Socket 6), the LDR Luminosity Sensor (Socket 2) and the PIR Presence Sensor (Socket 7). Detailed features of these sensors can be found in \cite{WaspEventsSB}. Table~\ref{table:consumption} summarizes the energy consumption involved in the Waspmotes operating as sensor nodes in our network.

\subsubsection{Software}

Four different applications can be deployed on the sensor nodes in our testbed: temperature monitoring, humidity monitoring, luminosity monitoring and presence detection. The first three ones monitor the corresponding variable of interest periodically at a configurable sensing interval, whereas the last one works by events that are generated when presence is detected. These applications have been developed using the Integrated Development Environment for Waspmote (Waspmote IDE) \cite{WaspmoteIDE}, which is compatible with the Arduino programming environment. Thus, Waspmotes  can only execute a given firmware at a time. Since our proposal aims to simultaneously deploy any subset of the available applications at a sensor node, the adopted solution has been to develop 15 different firmwares, one for each possible combination of the four considered applications. They are shown in table~\ref{table:process_cod}. For instance, firmware 1 just monitors temperature, while firmware 7 monitors luminosity, humidity and temperature. It is worth mentioning that even if only one firmware can be executed at a time, the remaining ones can be stored in the SD card and loaded when needed, as it will be later explained.

\begin{center}
\begin{table} [htpb]
\centering
\caption{Energy consumption values}
\begin{small}
	\centering
	 \adjustbox{max width=\textwidth}{
	\begin{tabular}{| r || c | c |}
 		\hline
 	
\textbf{Module} & \textbf{Mode} & \textbf{Consumption} \\ \hline \hline
	\multirow{2}{1.4cm}{\textbf{Waspmote}} & ON & 15 mA \\ 
\cline{2-3}
			& Sleep & 55 $\mu$A  \\ 
\hline
\multirow{4}{1.4cm}{\textbf{XBee ZigBee PRO}} & ON & 45.56 mA  \\ 
\cline{2-3}
& Sleep & 0.71 mA  \\ 
\cline{2-3}
& Sending & 105 mA  \\ 
\cline{2-3}
& Receiving & 50.46 mA  \\ 
\cline{2-3}
\hline
\multirow{6}{1.4cm}{\textbf{Events Sensor Board}} & Minimum (constant) & 3.6 $\mu$A  \\ 
\cline{2-3}
& Connector 2 + LDR S. & (32-64) + 0 $\mu$A  \\ 
\cline{2-3}
& Connector 5 + Temperature S. & 32 + 6 $\mu$A  \\ 
\cline{2-3}
& Connector 6 + Humidity S. & 32 + 180 $\mu$A  \\ 
\cline{2-3}
& Connector 7 + PIR S. & 0 + 100 $\mu$A  \\ 
\cline{2-3}
& Reading register ($<$20ms) & 150 $\mu$A  \\ 
\cline{2-3}
\hline
\multirow{4}{1.4cm}{\textbf{SD}} & ON & 0.14 mA  \\ 
\cline{2-3}
& Reading & 0.2 mA  \\ 
\cline{2-3}
& Writing & 0.2 mA  \\ 
\cline{2-3}
& OFF & 0 mA  \\ 
\cline{2-3}
\hline
			
\hline
    	\end{tabular}}
\end{small}
\label{table:consumption}
\end{table}
\end{center}

\begin{center}
\begin{table} [htpb]
\centering
\caption{Firmware codification }
\begin{small}
	\centering
	 \adjustbox{max width=\textwidth}{
	\begin{tabular}{| r || c | c | c | c |}
 		\hline
 	
\textbf{Firmware} & \textbf{PIR}  & \textbf{Luminosity}& \textbf{Humidity}& \textbf{Temperature}\\ \hline \hline
\textbf{1} &  &  &  & x   \\  \hline
\textbf{2} &  &  & x &    \\  \hline
\textbf{3} &  &  & x & x   \\  \hline
\textbf{4} &  & x &  &    \\  \hline
\textbf{5} &  & x &  & x   \\  \hline
\textbf{6} &  & x & x &    \\  \hline
\textbf{7} &  & x & x & x   \\  \hline
\textbf{8} & x &  &  &    \\  \hline
\textbf{9} & x &  &  & x   \\  \hline
\textbf{10} & x  &  & x &    \\  \hline
\textbf{11} & x &  & x & x   \\  \hline
\textbf{12} & x & x &  &    \\  \hline
\textbf{13} & x & x &  & x   \\  \hline
\textbf{14} & x & x & x &    \\  \hline
\textbf{15} & x & x & x & x   \\  \hline

\hline
			
\hline
    	\end{tabular}}
\end{small}
\label{table:process_cod}
\end{table}
\end{center} 

These firmwares have to perform two main functionalities: monitoring/sensing the parameters required by the applications (with configurable periodic sensing intervals or by events) and listening to the messages sent by the central controller (mainly firmware updates or changes in the applications sensing intervals). These two functionalities have been handled through alarms that allow waking up the Waspmotes from the low energy consumption state to perform the sensing and listening procedures, as it is explained next.

\begin{figure}[!t]
\centering
\includegraphics[width=3.1in]{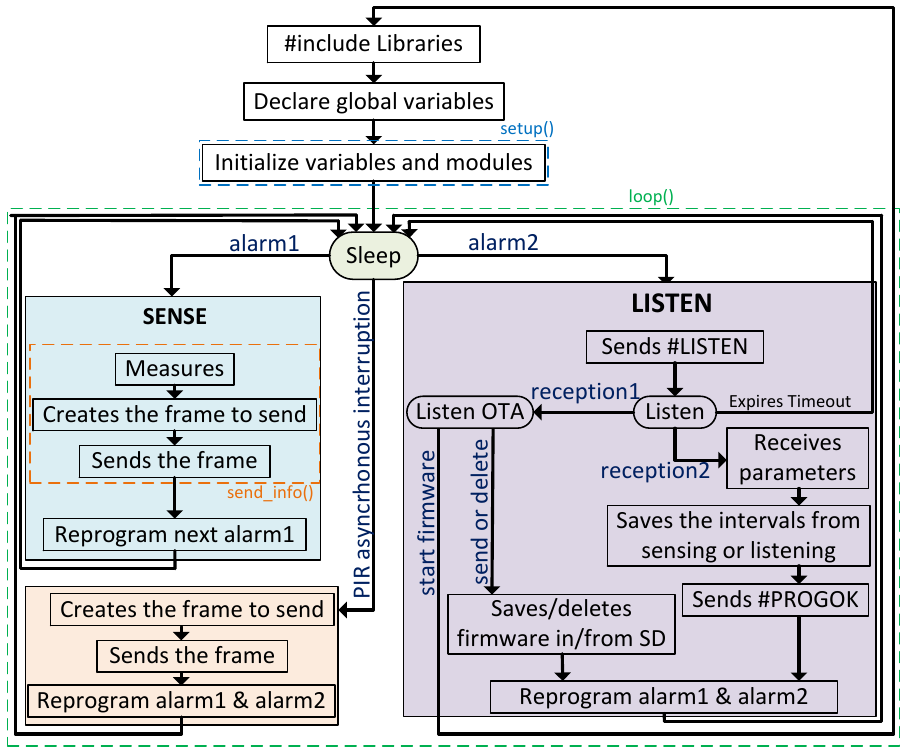}
\caption{Sensor node firmware schema}
\label{fig_sensor_schema}
\end{figure}

Figure~\ref{fig_sensor_schema} shows schematically the execution of these functionalities. Once the required libraries are included and variables are declared, the setup block configures the following modules: (i) XBee module for communications; (ii) RTC module for alarm programming; and (iii) Events Sensor Board for interpreting data from the sensors and transmitting them to the microcontroller. Once a device is connected to the network, it sends an information message to the central controller to inform about which applications are running and with which parameters. This message has the following format:

\begin{center}
\begin{scriptsize}
$|\#||INFO||\#||APP:x||\#||SENSOR:y||\#||LSTN:z|$
\end{scriptsize}
\end{center}

\noindent where \# is the delimiter, INFO is the frame identifier, APP:x is the firmware that it is running x $\in$ [1,...15], SENSOR:y is the sensing interval (in seconds) used by each application (SENSOR $\in$ [TEMP, HUM, LDR]; PIR is not included since it works by events), and LSTN:z is the listening interval (in minutes), with z $\in$ N. 

Finally, the alarms are initialized. \textit{Alarm1} is used for the sensing intervals and \textit{alarm2} is used to program the listening intervals, which allow checking with the controller for updates of the firmware. We set this interval to 1 minute.

Then, the device starts executing the loop() block, where the Waspmote remains in the Sleep mode  until any interruption happens. This low consumption mode allows configuring the modules that should or not sleep. This is necessary since, as stated in section \ref{System_communications}, the XBee module cannot be switched off in the routers, where applications can also be deployed.

PIR detection is activated with an asynchronous interruption when the sensor detects any movement, so it is originated by the Events Sensor Board and it is not programmable. Since this interruption is not predicted, to avoid missing any other interruption that may occur at time instants close to it, \textit{alarm1} and \textit{alarm2} are reprogrammed after this detection. When the \textit{alarm1} interruption happens (i.e., when it is time to measure the temperature, humidity, luminosity or a combination of them), data are sensed, encapsulated in a frame and sent to the central controller. This frame also includes the information about the remaining battery life in the sensor node, that will be used by the central controller to reallocate applications when needed. Finally, the next \textit{alarm1} is programmed.

As stated before, the central controller can send both firmware updates or changes in the application parameters. Firmware updates are performed using the OTA-Shell provided by Libelium \cite{OTA}, a functionality which allows the Over-The-Air Programming of Waspmotes through a multihop ZigBee network, which will be further explained in section \ref{architec_central}. 

 The XBee module distinguishes between the reception of standard and OTA frames: standard ones are directly processed and data is extracted, whereas for OTA frames, before the processing, it is checked which command contain. In order to process both type of frames using only one interruption, in our implementation the controller first sends a message indicating which kind of frame (standard or OTA) will send. 

So when the \textit{alarm2} interruption happens, the Waspmote sends a message to the central controller advising that it will start listening, thanks to the \#LISTEN command. After this, the Waspmote starts listening if any message from the central controller has been sent. If nothing is received, it goes to the sleep mode, whereas if it receives something, it first checks if it is an OTA or a standard frame to set the suitable reception mode. For OTA frames, if the \textit{$-send$} command is received, the device will save the firmware in the SD card, whereas if the \textit{$-delete\_program$} command is received, the device will access the SD storage and will delete the specified firmware. If the received command is a \textit{$-start\_new\_program$}, it will stop its execution to restart the device and execute the new firmware.

On the other hand, when a standard frame with application configuration parameters is received, the device will process them and it will send an acknowledgment message (\#PROGOK) to the central controller to confirm the update of the application parameters. Finally, both \textit{alarm1} and \textit{alarm2} are reprogrammed. The reprogramming of \textit{alarm1} is required because if the reception of a new firmware takes too much time, it could cause the lost of a preceding \textit{alarm1} and consequently that alarm would not be reprogrammed anymore. 

\begin{figure}[!t]
\centering
\includegraphics[width=3.1in]{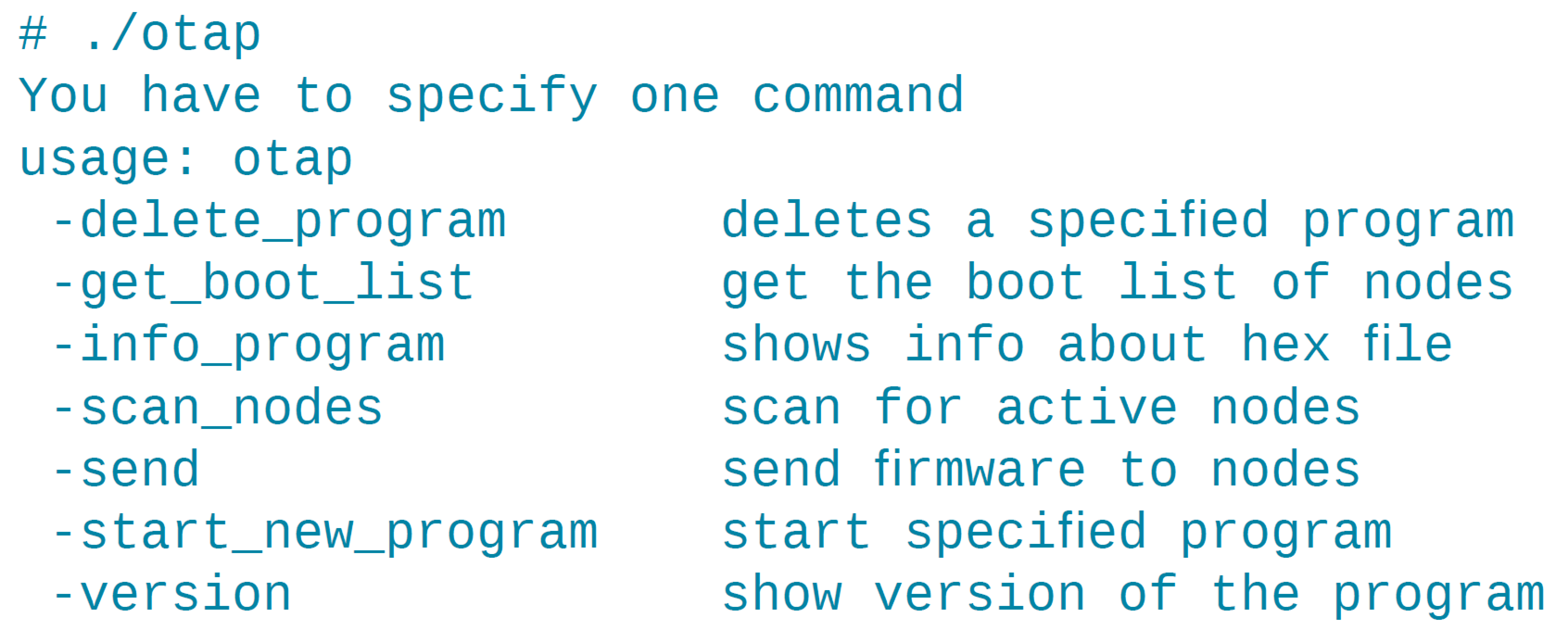}
\caption{Commands from the OTA-Shell}
\label{fig_otashell}
\end{figure}

\subsection{Central controller}
\label{architec_central}

The central controller has been implemented through a Java application deployed in a computer with a Java virtual machine. In order to provide ZigBee connectivity, an USB Waspmote Gateway from Libelium \cite{Libelium} with an XBee ZigBee-Pro S2 module has been used (see purple rectangle of central controller in Figure~\ref{fig_scenary}). The OTA-Shell specifications require configuring the serial communications between this Gateway and the XBee module to 38400 bps \cite{OTA}. The election of Java allows using the XBee Java Library, required to interact with the XBee module. Along with the Java application it is also used the aforementioned OTA-Shell provided by Libelium (Figure~\ref{fig_otashell}) to perform the OTA Programming of the Waspmotes. In addition, to manage all the collected data from the sensor nodes in the Java application, a database has been implemented with an XAMPP server managed with MySQL through PhpMyAdmin. Finally,  as a proof of concept, a simple graphic web interface hosted in the XAMPP server has been designed with PHP to show in real time the data from the nodes. 

\subsubsection{\textbf{Java application}}

The Java application is the main module of the central controller, where the system intelligence will reside. From the perspective of the SSN-IP that owns the infrastructure, this central controller should address the problem on how to optimally manage the application requests coming from different application providers to maximize the total revenue, deciding whether and when to admit an application request and how to reconfigure consequently the current physical resource assignment to applications, as we show in \cite{Delgado18}. To this purpose, parameters such as the activity time (the required lifetime of the application in the network) and a set of requirements to deploy the application can be considered in the decision process. These requirements could be the sensing area to be covered, the required processing and storage capabilities of the node(s) hosting the application, and the required communication bandwidth to deliver application data remotely across the physical sensor network. It is worth mentioning that in this work we focus on the implementation of the practical mechanisms to deploy multiple applications on the sensor nodes or reallocate them using OTA, rather than the decisions of where and when to perform these allocations, which we analyzed in \cite{Delgado18}.

\begin{figure}[!t]
\centering
\includegraphics[width=3.1in]{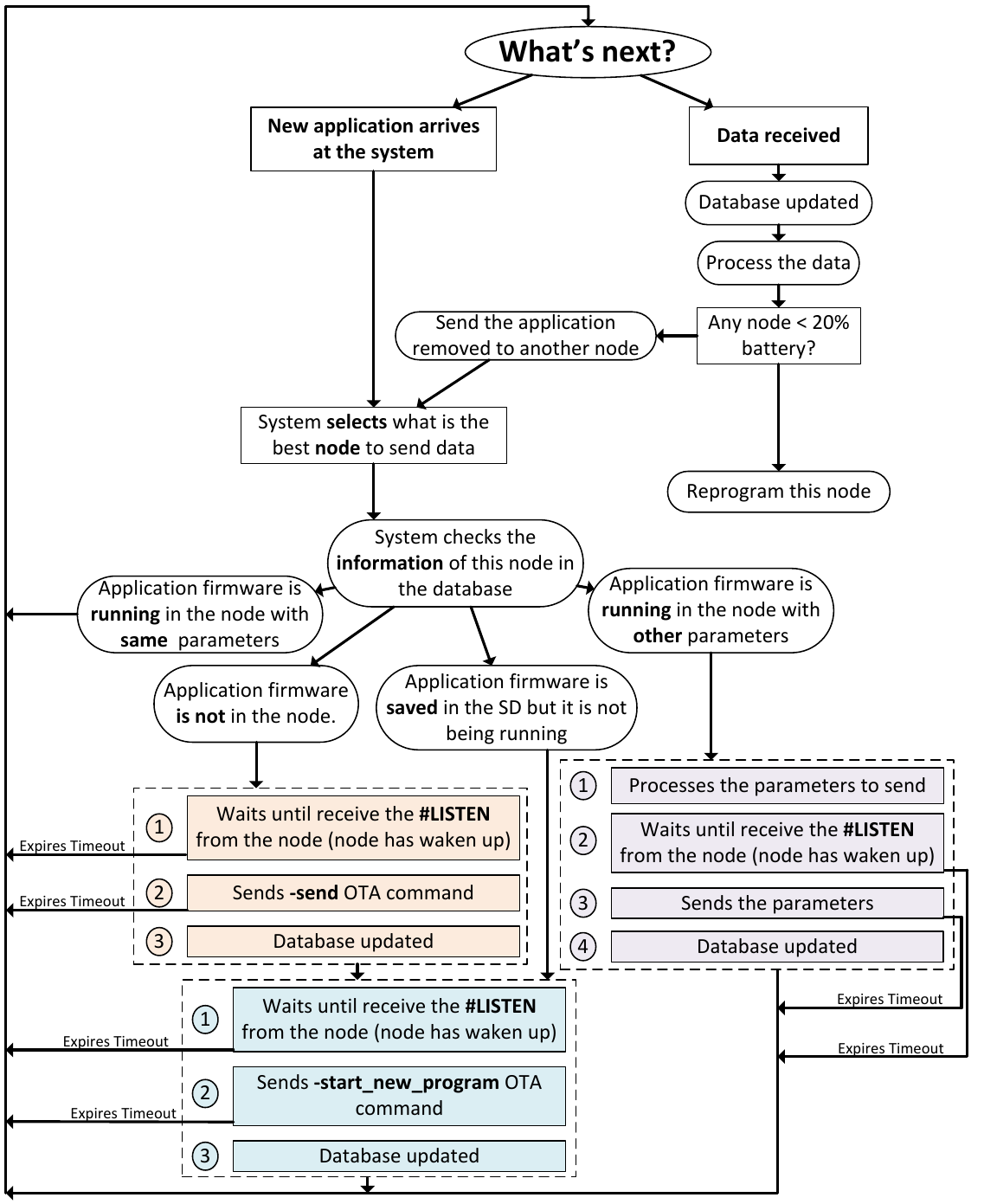}
\caption{Java application schema}
\label{fig_JavaSchema}
\end{figure}

Figure~\ref{fig_JavaSchema} shows the flow diagram of the Java application that the controller runs. When a new application arrives at the system, an algorithm should select the best node or nodes to allocate the application. As a simple example, we choose the node with the highest battery life. According to the status of the selected node, four different cases are possible:
\begin{enumerate}
\item Application is already running with the same sensing interval: Nothing is required. The current deployed application can provide the required data.
\item Application is running in the node with different sensing interval: A new configuration fulfilling both sensing intervals must be generated by the controller and sent to the node. As can be seen in the purple area of Figure~\ref{fig_JavaSchema}, the controller waits until it receives the \#LISTEN command from the node and then sends the new configuration frame to the node. When the acknowledgment is received, the database is updated.
\item Application firmware is saved in the SD card of the node but it is not running: The $-start\_new\_program$ OTA command is sent after receiving the \#LISTEN command. Once the acknowledgment is received, the database is updated. Note that this stored firmware must include the code of the active applications at that moment plus the new one. For example (as can be seen in Table \ref{table:process_cod}), if firmware 4 is running (luminosity) and a humidity application arrives to the system, firmware 6 must be activated.
\item Application firmware is not in the node: The $-send$ OTA command with the new firmware is sent after receiving the \#LISTEN command. When the acknowledgment is received, the database is updated. After this, it is neccesary to start this firmware, so the previous case (the blue one of the Figure~\ref{fig_JavaSchema}) is also executed.
\end{enumerate}

In addition, and also as a proof of concept to validate the application reallocation, the battery information included in all the frames from the sensor nodes (see section \ref{architec_sensor}) is compared to a predefined threshold. This threshold is set to 20\% because the manufacturer does not guarantee the correct reception and execution of the OTA-Shell comands under this value. Then, the central controller reprograms the node deleting the application with the highest energy consumption (i.e., starting a new firmware without this application, which should be also sent to the node if it is not already stored in the SD card) and sends it to another node with enough energy following the same steps as if it were a new one. 

Finally, we describe in the following example how the controller informs to a sensor node about the sensing intervals of its different applications.

\begin{figure}[hptb]
\centering
\includegraphics[width=3.1in]{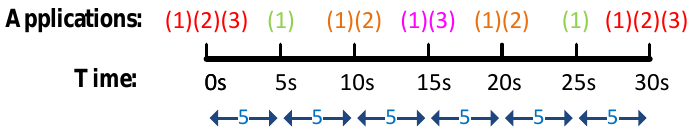}
\caption{Mechanism to process the intervals}
\label{fig_intervalprocess}
\end{figure}

Let us assume we have to sense temperature (1) every 5 seconds, humidity (2) every 10 seconds and luminosity (3) every 15 seconds. This combination of applications corresponds to firmware 7. Figure~\ref{fig_intervalprocess} shows the time instants in which each application will be sensed. Since every 30 seconds the same structure will be repeated, the following data frame is sent:

\begin{center}
\begin{scriptsize}

$|$2$||$\textcolor{blue}{$<$5$><$5$><$5$><$5$><$5$><$5$>$}$||$-$||$\textcolor{red}{$<$7$>$}\textcolor{green}{$<$1$>$}\textcolor{orange}{$<$3$>$}\textcolor{magenta}{$<$5$>$}\textcolor{orange}{$<$3$>$}\textcolor{green}{$<$1$>$}\textcolor{red}{$<$7$>$}$||$

\end{scriptsize}
\end{center}

\noindent where the first $|$2$|$ means that this is a standard data frame with sensing intervals, $|<$5$><$5$><$5$><$5$><$5$><$5$>~|$ are the time intervals between two sensings, and $|<$7$><$1$><$3$><$5$><$3$><$1$><$7$>|$ represents the indexes of the firmwares whose applications must be sensed at each time.

\subsubsection{\textbf{Database}}

\begin{figure}[!t]
\centering
\includegraphics[width=2.8in]{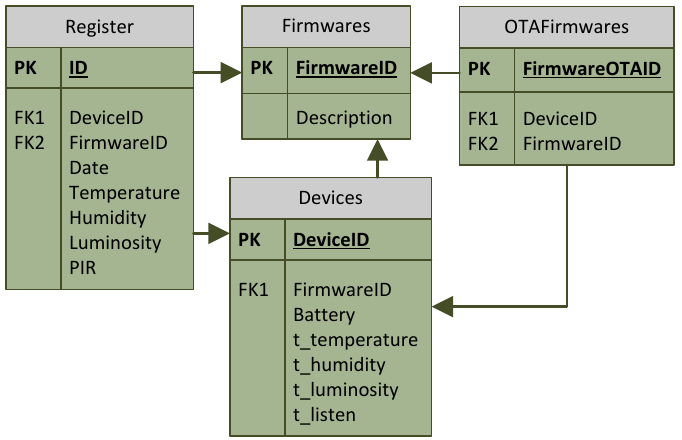}
\caption{Entity relationship database diagram}
\label{fig_database}
\end{figure}

To manage the database MySQL through PhpMyAdmin are used, integrated in XAMPP, from where new databases and their tables can be created and registers can be inserted with SQL sentences. Figure~\ref{fig_database} shows the entity relationship database diagram that will help to understand its functionality. The database is formed by four tables: \emph{Register, Devices, Firmwares} and \emph{OTAFirmwares}, where primary and foreign keys have been used. Primary ones are identified with \textit{PK}, while the foreign ones are identified with \textit{FKx}. The register table has all the registers that the controller receives from the nodes. Its primary key is the ID, whose value is increased everytime a new data is received. Foreigns keys are \emph{DeviceID} and \emph{FirmwareID}, which allow to access to the device registers and the firmware registers. The \emph{Devices} table has as primary key the \emph{DeviceID} attribute, which is the identifier of each node. It contains the information of the nodes: battery, listening time and sensing intervals of the applications. The firmware table contains the information of the firmwares that can be executed in the nodes, and \emph{FirmwareID} is its primary key. \emph{OTAFirmwares} saves all the FirmwareIDs stored in the SD memory card of the network devices. Its primary key is \emph{FirmwareOTAID}, and its foreigns keys are \emph{DeviceID} and \emph{FirmwareID}.

\subsubsection{\textbf{Graphic web interface}}

\begin{figure}[!t]
\centering
\includegraphics[width=2.8in]{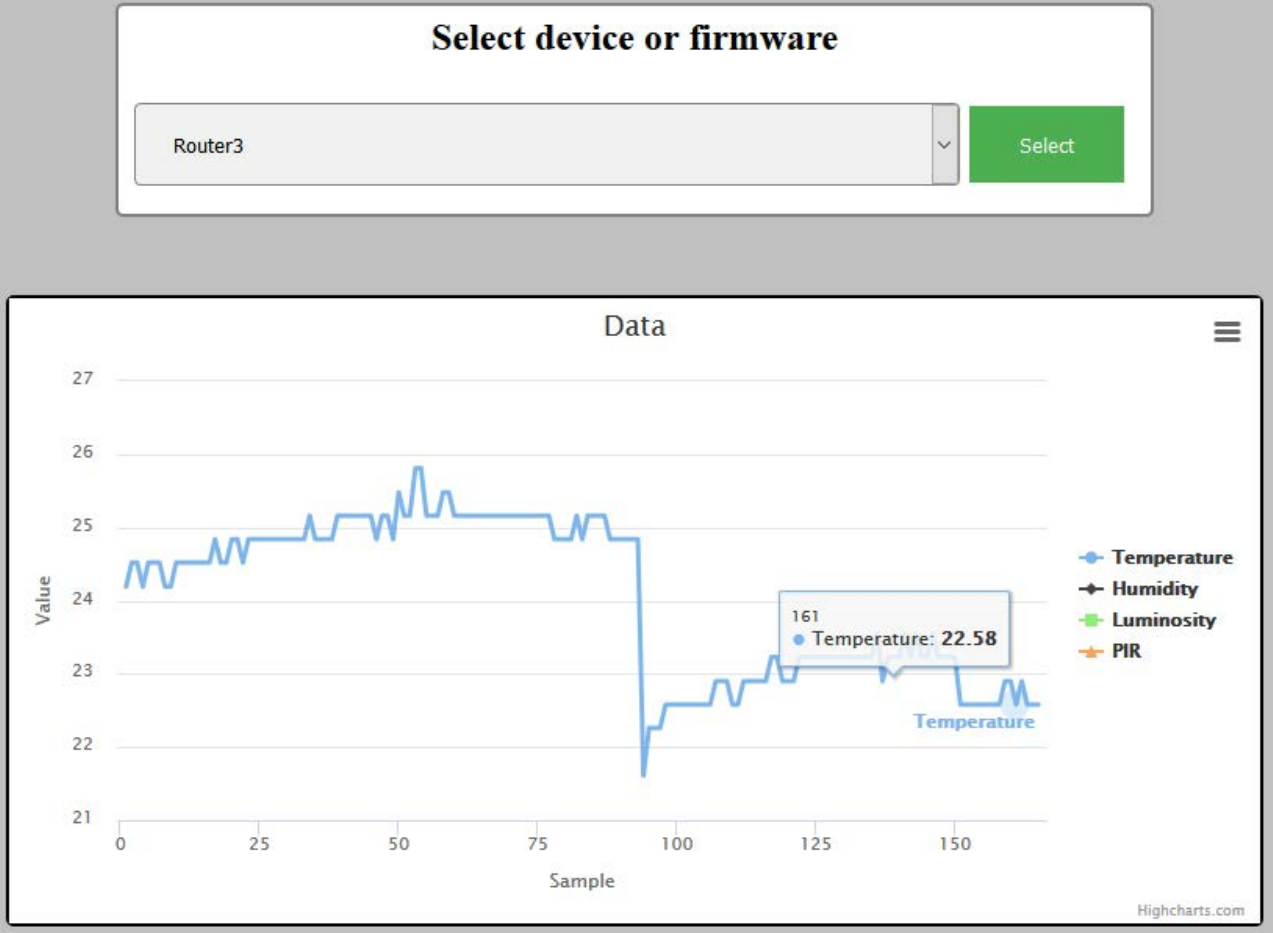}
\caption{Graphic web interface}
\label{fig_web}
\end{figure}

A simple graphic web interface has been designed to monitor all the data sensed by the nodes in real time. It has been designed with PHP and is hosted in the XAMPP server. The Highcarts library has been used for the graphic design. The stored parameters can be shown by device or by application. Figure~\ref{fig_web} shows a screenshot of the graphic web interface monitoring several applications. 

\section{Experimental Results}
\label{sec:results}

In this section, we evaluate the performance of the system implementation. Two main aspects have been analysed: the time required to perform a firmware update and the energy consumption of the sensor nodes.

\subsection{Firmware update}

Firmware update is one of the key elements to perform the adaptive resource allocation aimed with this implementation. Two possible scenarios can arise: If the firmware to be updated is already saved in the SD card of the sensor node, it only has to be started with the \textit{$-start\_new\_program$} command. However, if the firmware is not in the sensor node yet, it has to be sent from the central controller with the \textit{$-send$} command and then run with the \textit{$-start\_new\_program$} command.

   Table~\ref{table:rates} shows the results in terms of number of frames, firmware size, ZigBee data bytes sent (81 bytes per 802.15.4 frame), required time, and effective ZigBee data rate when updating firmware from the coordinator to a 1-hop router (Figure \ref{fig_schemas}(a)) for different application sets. 
   
\begin{table*}[t!]\footnotesize 
	
	\renewcommand{\arraystretch}{1.3}
	
\begin{center}

\centering
\caption{Firmware update results for different application sets}
\begin{small}
	\centering
	 \adjustbox{max width=\textwidth}{
	\begin{tabular}{| r | r || c | c | c | c | c |}
 		\hline
 	
\textbf{Update} & \textbf{Firmware: Applications} & \textbf{Frames}  & \textbf{Firmware size} & \textbf{Bytes sent} & \textbf{Time}& \textbf{Rate}\\ \hline \hline

	\multirow{4}{1.4cm}{\textbf{\textit{-send} firmware}} & 1: Temperature  & 982 & 74080 B & 79542 & 115 s & 5.53 kbps \\ 
\cline{2-7}
& 3: Temperature and Humidity  & 1058 & 79704 B & 85698 & 129 s & 5.31 kbps \\ 
\cline{2-7}
& 7: Temperature, Humidity and Luminosity & 1061  & 79754 B & 85941 & 132 s & 5.2 kbps \\ 
\cline{2-7}
& 15: Temperature, Humidity, Luminosity and Presence  & 1071  & 80506 B & 86751 & 134 s & 5.17 kbps \\ 
\hline
	\multirow{4}{1.4cm}{\textbf{\textit{-start new program}}} & 1: Temperature  & 1 & - & 19 & 0.011 s & 13.17 kbps \\ 
\cline{2-7}
& 3: Temperature and Humidity  & 1 & - & 19 & 0.01 s & 14.52 kbps \\ 
\cline{2-7}
& 7: Temperature, Humidity and Luminosity  & 1 & -  & 19 &0.0135 s & 11.23 kbps \\ 
\cline{2-7}
& 15: Temperature, Humidity, Luminosity and Presence & 1 & - & 19 & 0.009 s & 15.24 kbps \\ 
\hline
\hline
			
\hline
    	\end{tabular}}
\end{small}
\label{table:rates}

\end{center} 
\end{table*} 
 
As expected, the time required to activate an already stored firmware is much lower than the one required to send it, since only one frame has to be sent. A 2GB SD card allows to store more than 16000 firmwares, which are limited to 128 KB (flash memory of Waspmote, see section~\ref{hardware}). In our example scenario, with 15 firmwares, all of them can be preloaded in the SD card.

It is also worth noting that when the firmware is sent, not only the amount of information to be sent is higher, but the effective transmission rate is lower. This result directly depends on the OTA-Shell implementation from Libelium, which is used to send the firmwares to the sensor nodes. As for the data rate obtained in the other case, in the XBee documentation \cite{XBeeProUser} it is stated that empirical results show that the maximum achievable ZigBee data rates at one hop router to router links with security disabled is 35 kbps when the serial communications between XBee module and device are set to 115200 bps. Note that in our scenario, serial communications between XBee module and central controller must be set to 38400 bps.  

\begin{figure}[!t]
\centering
\includegraphics[width=2.5in]{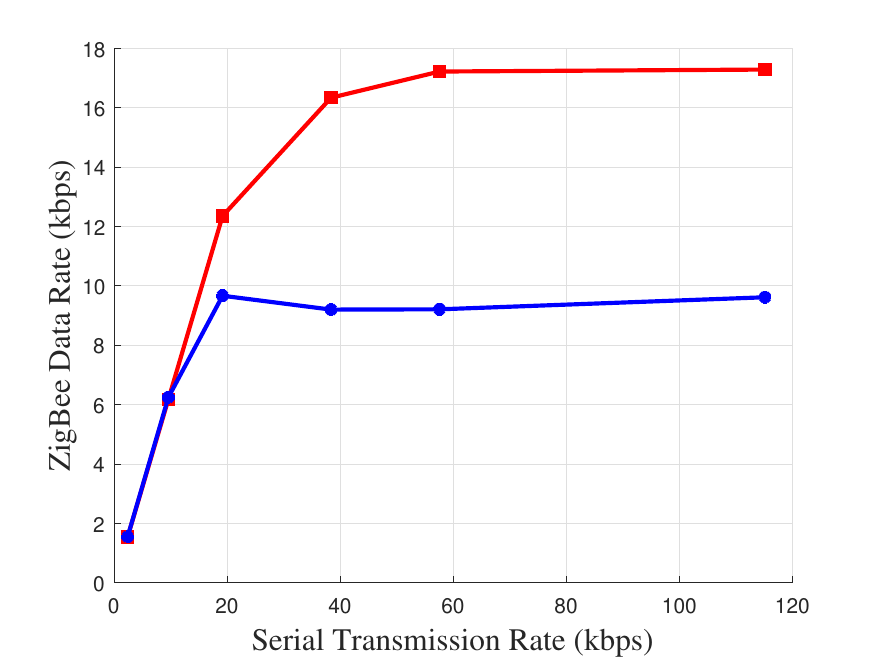}
\caption{Transmission rate from the controller}
\label{fig_grafica}
\end{figure}

We have analyzed the maximum achievable data rate in our scenario both for 1-hop communications between the coordinator and a router and 2-hop communications between the coordinator and an end device (Figure \ref{fig_schemas}(a)) varying the serial communications rate between XBee module and coordinator. Figure~\ref{fig_grafica} shows the obtained results, slightly lower than the maximum values shown in \cite{XBeeProUser}, but confirming the data rates shown in table~\ref{table:rates}. 

\subsection{Energy consumption}

\begin{figure}[!t]
\centering
\includegraphics[width=3.1in]{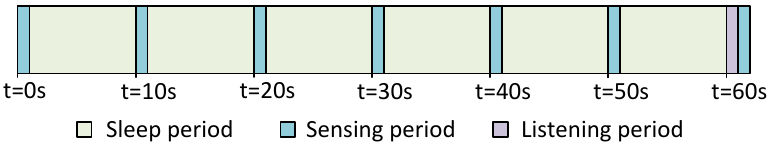}
\caption{Temporal diagram of a Waspmote program, sensing every 10s and listening every 1m}
\label{fig_temporalDiagram}
\end{figure}

In this section, as an example to check the expected energy consumption in the sensor nodes, we have analyzed the case of a router in an extreme situation with very high energy consumption. The router is running a temperature application with a sensing interval of 10 s and it is listening to the coordinator 1 s every 60 s (as Figure~\ref{fig_temporalDiagram} shows).  

Eq. (\ref{eq_1}) shows its lifetime where $EC$ is the total energy consumption, which is defined in eq.~(\ref{eq_2}), and 6600 mAh is the battery capacity of the devices:

\begin{equation}
lifetime =  \frac{6600 mAh}{EC},
\label{eq_1}
\end{equation}

\begin{equation}
EC = SLPEC * t_{SLPEC} + SEC * t_{SEC} + LEC * t_{LEC},
\label{eq_2}
\end{equation}

\noindent where $SLPEC$ is the sleep energy consumption, $SEC$ is the sensing energy consumption, $LEC$ is the listening energy consumption and $t_{SLPEC}$, $t_{SEC}$ and $t_{LEC}$ are the corresponding fractions of time at each state.

In the sleep period, the Waspmote is in Sleep mode (0.055 mA), the XBee module is always on, since it is acting as a router (45.56 mA), the Events Sensor Board consumes at its minimum (0.0036 mA) and the SD is always on (0.14 mA). $t_{SLPEC}$ in this case is 0.9807. 

$SEC$ corresponds to the energy consumption related to the sensing itself and the sending of the sensed data when the Waspmote and its SD are always on (15mA and 0.14 mA, respectively). In this case, we distinguish between the sensing time, the ZigBee packet sending time and the ZigBee ACK receiving time. In the first case, the Xbee module is on (45.56mA) and the Events Sensor Board is sensing temperature (0.1916 mA, which also takes into account the reading register), which takes less than 20ms according to the specifications of the manufacturer \cite{WaspEventsSB} (fraction of time for sensing is 0.002). Secondly, the XBee is sending (105 mA) and the Events Sensor Board is in its minimum (0.0036 mA). The 802.15.4 temperature data frame has 102 bytes at PHY level, (around 3.3 ms, fraction of time for sending is 0.00033). Thirdly, the XBee will be receiving (50.46 mA) and the Events Sensor Board will be in its minimum again (0.0036 mA). The 802.15.4 frame for the ZigBee ACK frame has 49 bytes at PHY level (around 1.6 ms, fraction of time for receiving of 0.00016). 

Finally, for the listening intervals, we have to consider that the node first sends the \#LISTEN command (and the corresponding ZigBee ACK from the coordinator is received), and then it waits 1 s for receiving messages from the central controller (fraction of time of 0.01667). Note that in this case no message from the central controller is sent. The Waspmote is on (15 mA), the Events Sensor Board is in its minimum value (0.0036 mA) and the SD is on (0.14mA). First, the XBee is sending (105mA) and since the 802.15.4 frame for the \#LISTEN command has a length of 87 bytes, the transmission takes around 2.8 ms (fraction of time of 0.000046). The 802.15.4 frame for the ZigBee ACK frame has 49 bytes at PHY level (around 1.6 ms, fraction of time for receiving of 0.000026). 

With these parameters, we theoretically estimate a lifetime of 5.96 days. According to the experimental results, obtained measurements have shown that in 7 hours running the temperature application the battery dropped 11\%. From this result, we could estimate a battery duration of 2.65 days, which is in the same magnitude order. It must be noted that this reduced lifetime is due to the excesively short intervals choosen for the sensing and the listening and because calculations have been made at a router. Since ZigBee routers must have the XBee module active all the time, they are usually plugged into the grid or have an external energy source. In fact, the theoretical estimation of the lifetime of an end device (where the XBee module can be in sleep mode during the Waspmote sleep periods) running the same firmware, is 129.58 days.

\section{Conclusion}
\label{sec:conclusions}
In this paper we have presented the implementation of a system architecture for wireless SSNs that allows dynamically allocating applications arriving to the system or reallocating already deployed applications through over-the-air programming of the sensor nodes, which support multiple concurrent applications. Experimental results have shown the viability of the implementation. It is worth noting that storing in the sensor nodes all the possible deployed firmwares highly improves the performance of the over-the-air programming, so the firmwares should be either preloaded before the deployment or stored after the first update. As future research lines, the deployment of a larger network, where some of our previously proposed resource allocation algorithms could be tested and the comparison with other schemes is foreseen. 


\section*{Acknowledgment}
This work has been supported by the Spanish Government through the grant TEC2014-52969-R from the Ministerio de Ciencia e Innovaci\'on (MICINN), Gobierno de Arag\'on (research group T98), the European Social Fund (ESF) and Centro Universitario de la Defensa through project CUD2016-17.



%

\footnotesize
\bibliographystyle{IEEEtran}
\bibliography{./IEEEabrv,./bare_conf}

\end{document}